\documentclass[10pt]{article}

\usepackage[a4paper,twoside, left=4cm, width = 12cm, top = 5.2cm , bottom = 5.8cm, includehead]{geometry}
\usepackage{titlesec}
\titleformat{\section}
{\normalfont\rmfamily\Large}
{\thesection}{1em}{}
\titleformat{\subsection}
{\normalfont\rmfamily\large}
{\thesubsection}{1em}{}

\usepackage{fancyhdr}
\let\OLDthebibliography\thebibliography
\renewcommand\thebibliography[1]{
	\OLDthebibliography{#1}
	\setlength{\parskip}{0pt}
	\setlength{\itemsep}{3pt plus 0.3ex}
}

\usepackage[utf8]{inputenc}
\usepackage[english, german]{babel}
\usepackage{setspace}

\usepackage{float}
\usepackage[T1]{fontenc}
\usepackage{graphicx}
\usepackage[varg]{txfonts}

\normalsize

\begin{document}

\selectlanguage{english}
\setcounter{page}{0}

\begin{figure}[H]
	\begin{center}
		\resizebox{9cm}{!}
		{\includegraphics{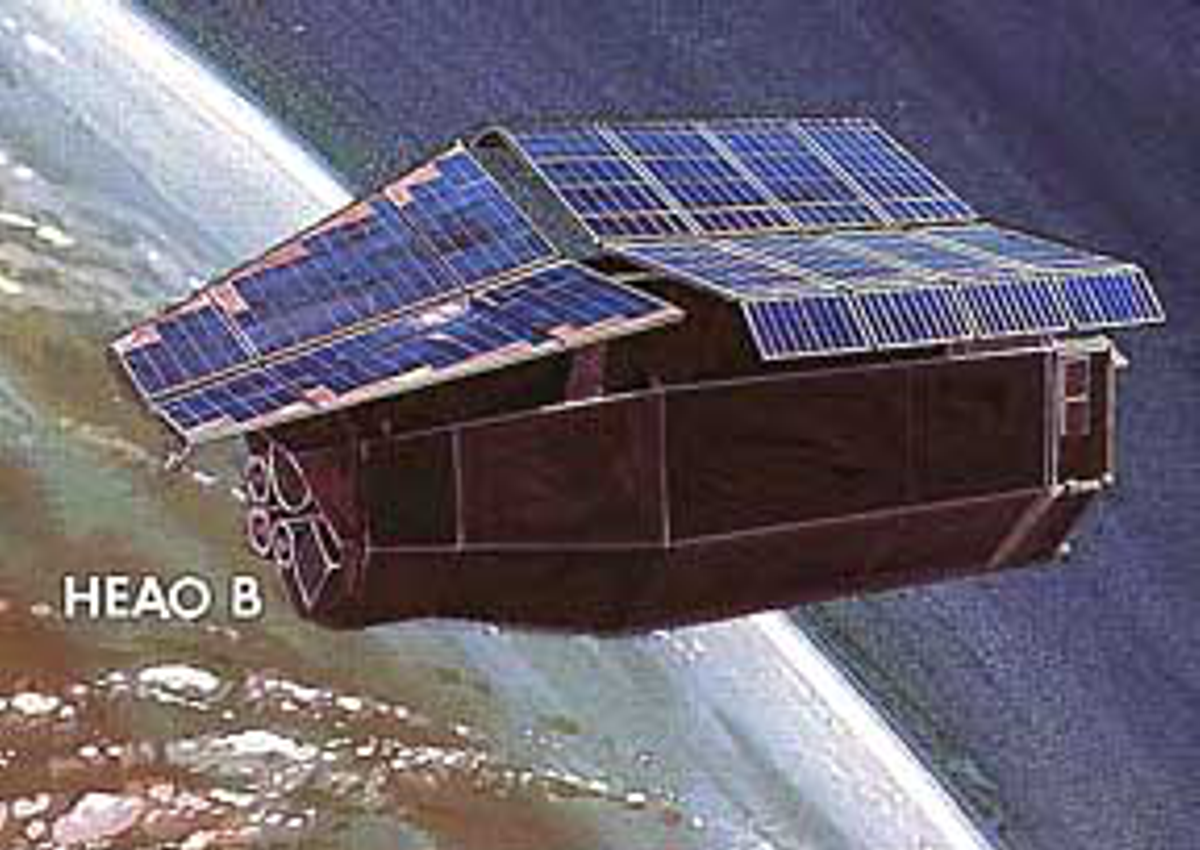}}
		
		\vspace{3mm}		
		
		\resizebox{5.2cm}{!}
		{\includegraphics{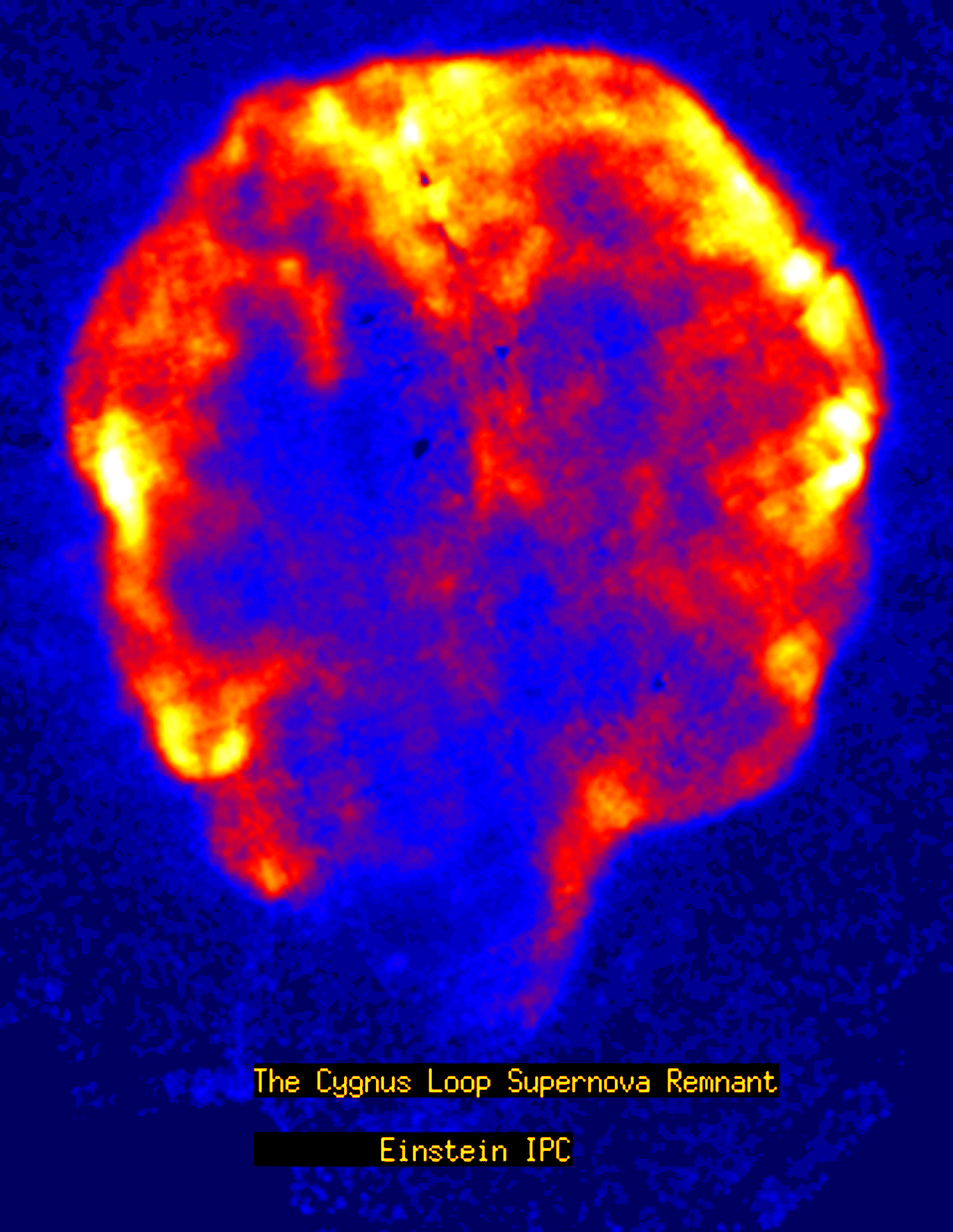}}	
		\label{Fig_HEAO_b}
	
		
	\caption{\ }
	\begin{minipage}{12.5cm}
		\centering
			Upper figure: The first imaging X--ray telescope, NASA's Einstein Observatory (HEAO--2), was launched on November, 12th 1978, 26 years after the proposal of a grazing--incidence imaging optics for X--rays by Hans Wolter. (Image NASA)

	Lower figure: X--ray image of the Cygnus Loop supernova remnant, taken with the Einstein Observatory (Image NASA)
	\end{minipage}

	\end{center}
\end{figure}


\pagebreak
\thispagestyle{empty}

\noindent
\LARGE{Hans Wolter -- a pioneer of applied optics}

\vspace{1cm}

\noindent
\Large{\textit{Andreas Schrimpf (Marburg)}
	

\normalsize


\vspace{1.5cm}

\noindent
\Large{Abstract}

\vspace{5mm}

\normalsize
\noindent
	Applied optics was one of the major topics Hans Walter was engaged in during his scientific life. He contributed to the understanding of optical properties of thin films, which could be used to design coating layers to improve the properties of optical and other surfaces. He developed the theoretical description of the basic principles of phase-contrast, schlieren and interference optics applied to enhance low contrast details and to increase the resolution in studies of biological samples. And last, but not least, Hans Wolter proposed an optical system of two grazing--incidence mirrors for use in an X--ray imaging microscope. A microscope using such an optics never was put into practice, but the optical design turned out to be well suited for telescopes.

\vspace{5mm}

\noindent
\Large{Zusammenfassung}

\vspace{5mm}

\normalsize
\noindent
	Die angewandte Optik war eines der wichtigsten Fachgebiete, mit denen sich Hans Wolter in seinem wissenschaftlichen Schaffen auseinander setzte. Seine ersten Beiträ\-ge beschäftigen sich mit den optischen Eigenschaften dünner Filme, die man z.B.\ zur Vergütung oder als Verstärkerfolien einsetzten konnte. Er hatte maßgeblichen Anteil an der Entwicklung von Grundlagen und Anwendungen von Phasen\-kontrast-, Schlieren- und Interferenzoptiken, die vor allem zur Kontrastverstärkung und zur Erhöhung der Auflösung in der Mikroskopie biologischer Proben eingeetzt wurden und werden. Und, wohl am bekanntesten ist Wolter für die Idee und theoretische Beschreibung einer abbildenden Röntgenoptik, die er für die Mikroskopie nutzen wollte. Sein Entwurf stellte sich als überlegenes Design für den Einsatz in Röntgenteleskopen heraus.
	

\pagebreak

\section{Biographical and personal remarks}
Hans Wolter (1911 -- 1978) was born and grew up in Dramburg (Western Pomerania). His interests in theoretical and applied physics attracted attention already at high school. In the certificate of qualification for university matriculation his school achievements were commented: ''\emph{Mathematik: Seine Leistungen gingen weit über das Ziel der Klasse hinaus: Sehr gut. Physik: Er zeigt auffallendes Verständnis für die mathematische Behandlung physikalischer Fragen. In der Optik besitzt er recht gute Kenntnisse: Sehr gut.}'' [\cite{Kiel1947}]. After school in 1929 Wolter went to Tübingen and later to Kiel for studying physics, mathematics and chemistry. He finished in 1933 with a state examination as a teacher. While working as a teacher he continued his research at Christian--Albrecht--Universität in Kiel, supervised by Albrecht Unsöld, and received his PhD on November, 7th, 1936.

He had to join the German navy from 1939 to 1945. After World War II Wolter run a private laboratory until in 1947 he became a scientific assistant at the university in Kiel. Wolter habilitated in 1949 and was appointed a ''Diätendozentur`` (assistant professorship) in 1952. 
In late 1954 Wolter was offered the position as a professor for applied physics at the Philipps--Universität Marburg, which he accepted. He moved to Marburg in May 1955, where he established the Institute of Applied Physics, which he led to a world wide excellent reputation. 

Wolter was a member of the scientific advisory council of several international journals, joint editor of two book series, member of the German Society of Applied Optics (DGaO) and a fellow of the Optical Society of America (OSA)
[\cite{Blume1979}].

Hans Walter was married and had four children, born between 1943 and 1949.
He retired in 1977 and died at the age of 67 on 17th August 1978.

Hans Walter was known to be a very accurate person. In his office every item had to be at the correct place. Once, some tools could not be found in the laboratories. He set up a document, which he passed to the lawyers of the university for checking correctness and legitimacy, which contained rules of action and responsibility in his labs [\cite{Wolter1966}]. Finally, he was known as a very committed and good teacher [\cite{Thomas2015}] He developed a series of lectures about applied physics and he stood for a good education in physics for students of medicine and life sciences.
	
\begin{figure}[t]
	\begin{center}
		\includegraphics[width=7.1cm]{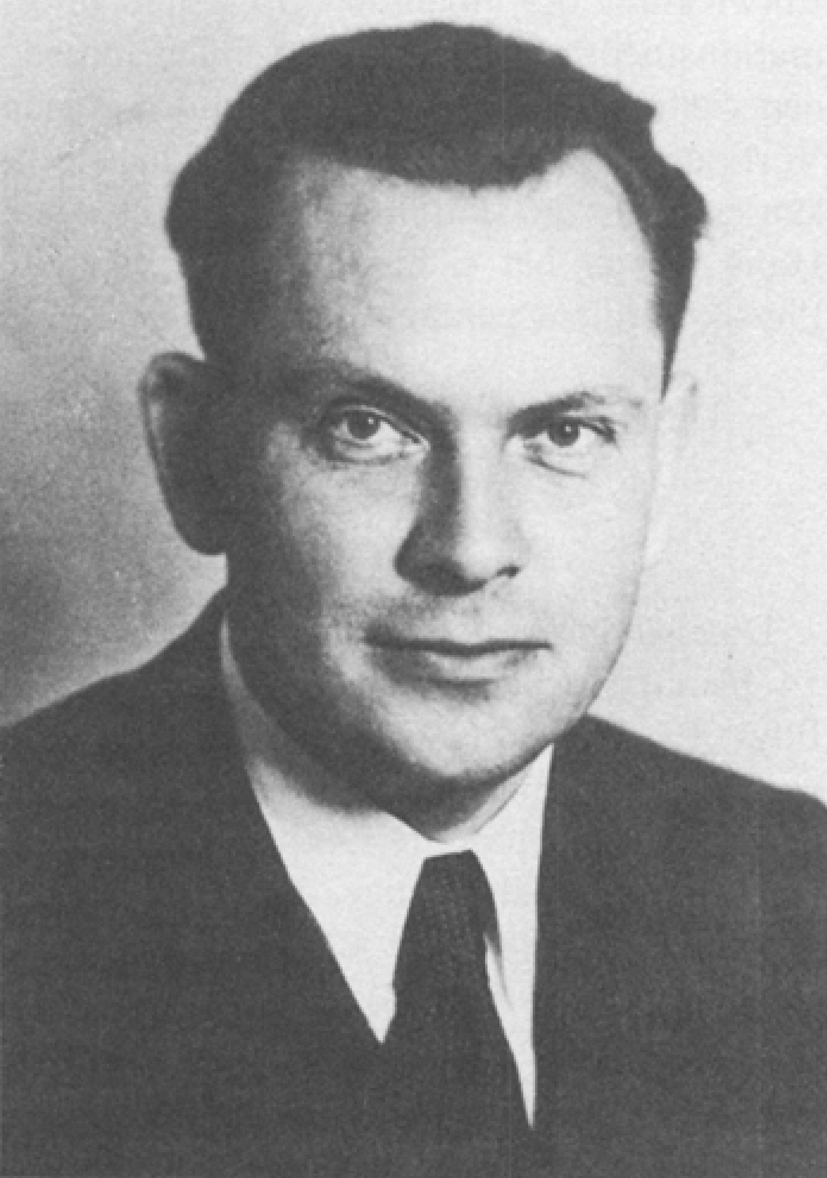}
		\caption{ }
		\begin{minipage}{8cm}
			\centering
			Hans Wolter (1911 - 1978)\\
			
			[\cite{Gradmann1979}]\\

	\end{minipage}
\end{center}
\end{figure}

\section{Hans Wolter in Kiel }
After finishing his studies at the university in Kiel, Wolter started to work at school as a teacher in probation. During his time at the university he met Albrecht Unsöld (1905 -- 1995), who offered him to supervise a PhD thesis in theoretical optics, a study of the optical properties of thin metal films [\cite{Wolter1937}]. When finishing his thesis, Wolter was at the age of 25 and Unsöld 31, a very inspiring coincidence [\cite{Gradmann2015}]. From 1937 to 1939 Wolter continued his research on theoretical optics [\cite{Wolter1939}], while working as a teacher at a local school. He used a more general approach of wave optics, that could and still can be used to describe many phenomena of the optics of thin films.

At the beginning of World War II the navy became interested in his research. He had to join the 
''Nachrichtenmittelversuchskommando`` of the German navy from 1939 to 1945. There he investigated the use of wideband and underwater antennas for communication, resulting in three patents  [\cite{Wolter1947}]. 

\begin{sloppypar}
After World War II Wolter was not allowed to continue his employment as a teacher in a public school. Living in Neustadt (Holstein) he founded a private laboratory. The production of  microscopical samples caused him to start developing different techniques to visualize non-absorbing objects with microscopes [\cite{Lochte1954}].
\end{sloppypar}

In 1947 Walter Lochte-Holtgreven (1903 -- 1987) offered Wolter a position as scientific assistant at the university in Kiel, where he received his venia legendi in 1949. The following years at Christian--Albrecht--Universität Kiel can be rated as the one of the most productive scientific time of Wolter's life. In 1950 and 1951 Wolter published 14 papers on phase-contrast, schlieren and interference optics to improve the investigation of microscopic biological samples (see section \ref{sec_schlieren}). In 1952 Wolter published his famous two papers about the imaging X--ray optics (see section \ref{sec_x-ray}). 

To enhance the visibility of structures in biological samples Wolter applied 2-dimensional color marking in his schlieren optics. To use the colors as source of physical information, Wolter developed a physical theory of colors resulting in a system of standard colors  [\cite{Wolter1950c}].

Wolter became known as expert for applied optics. In 1950 the ''Optische Werke Ernst Leitz'' asked Wolter for a cooperation about new techniques for microscopes. Wolter agreed and finally sold a patent to them  [\cite{Lochte1950}]. 1952 he gave an invited talk at the annual meeting of German physicists about microscopy using X--rays and visible light, and in 1953 the Springer publishing company asked him for a review article about optics of thin films and phase-contrast microscopy for the new edition of the ''Handbuch der Physik''. Wolter wrote two review articles, which were published in 1956 [\cite{Wolter1956a}, \cite{Wolter1956b}].

\section{Hans Wolter in Marburg}

After World War II there existed three physics institutes in the Philosophical Department of the Phillips-Universität Marburg: Erich Hückel (1896 -- 1980) was the head of the Institute of Theoretical Physics, Wilhelm Walcher (1910 -- 2005) was the head of the Institute of Experimental Physics and Siegfried Flügge (1912 -- 1997) was heading the Institute of Structure of Matter. Walcher pleaded for a chair of applied physics, but in 1947 and 1948 two search procedures failed. In 1947 Walcher was appointed acting professor of applied physics. In 1952 a third search procedure was launched and letters were sent to several institutes to ask for recommendation of suitable scientists. It was Walther Gerlach (1889 -- 1979) who suggested to consider Hans Wolter [\cite{Gerlach1952}]. The final list of candidates was sent to the state administration in Wiesbaden in 1954 including a letter of reference from Walter Lochte--Holtgreven  [\cite{Lochte1954}]. At May, 1st 1955 Hans Walter became adjunct professor of applied physics at the Philipps-Universität Marburg and in 1959 full professor, after refusing an offer of the Universität des Saarlandes [\cite{Uni_MR_1959}].

In 1954 Walcher had already applied for a new building for the Institute of Applied Physics, close to the two buildings of the other physics institutes at Renthof 5 and Renthof 6. Wolter took up these plans, modified the specifications to his needs and after a long struggle to move the forestry office and to get the estate finally in spring 1960 the Institute of Applied Physics could move into the new building at Renthof 7 [\cite{Uni_MR_1960}]. Interestingly the stories of the building were painted in colors according to Wolter's theory of colors  [\cite{Gradmann2015}]; unfortunately this coloring has not been preserved.

In Marburg Wolter's research topics changed. Ulrich Gradmann was one of the first scientific assistants joining the group. Gradmann started his own research on structure and magnetism of oligatomic films. Gradmann became professor in 1972 and left Marburg in 1977. Siegfried Blume joined the group in 1960. He picked up Wolter's research about properties of antennas. By the end of 1958 Wolter got a measuring station in a distance of about 1.5 km from the institute to study the far field, and in 1959 a second measuring state about 100 m from the institute was set up for studying the near field properties of antennas [\cite{Wolter1959}]. Blume became professor in 1972 and left the institute in 1977. Wolter himself was mainly working on topics of the information theory in optics and communication systems with basic contributions to sampling theorems. And, he continuously was interested in applications of his results in medicine and biology [\cite{Wolter1976}].

\section{Selected topics of Applied Optics}

The three most cited topics of Wolter's work in decreasing order are the X--ray optics, the studies of optics of thin metal films and the minimum beam marking in schlieren photography. In the following the basic ideas of two of them, the minimum beam marking and the X--ray optics, well be presented in more details.

\subsection{Sub-diffraction resolution in optical systems}
\label{sec_schlieren}

Many microscopic biological samples are composed of non-absorbing matter. However, the optical density could be different due to different composition, different density of the material. So imaging light rays passing through more dense or less dense parts of an object would be refracted into different directions. Placing a blade into the optical path cuts out all the rays starting from a chosen angle of refraction. Thus, in the image plane there appears a dark and a light part of the image with the cross section showing the parts of the sample with the marked refraction angle (see right upper part of fig. \ref{fig_minimum_beam_marking}). This technique was developed by the German chemist and physicist August Toepler in the 19th century.

During his investigations of optics for non-absorbing microscopic samples, Wolter realized that the resolution of a maximum in a 2-dimensional interference field is limited by approximately one wave length (diffraction limit), while a minimum as a zero point is infinitesimally narrow! He replaced the blade by a half-wave plate, covering half of the beam, and got a tremendous improvement in resolution, well below the diffraction limit of light [\cite{Wolter1950a}, \cite{Wolter1950b}]. He called this method the \emph{minimum beam marking} (Figure \ref{fig_minimum_beam_marking}).

\

\begin{figure}[H]
	\begin{center}
		\label{fig_minimum_beam_marking}
		\includegraphics[width=0.48\linewidth]{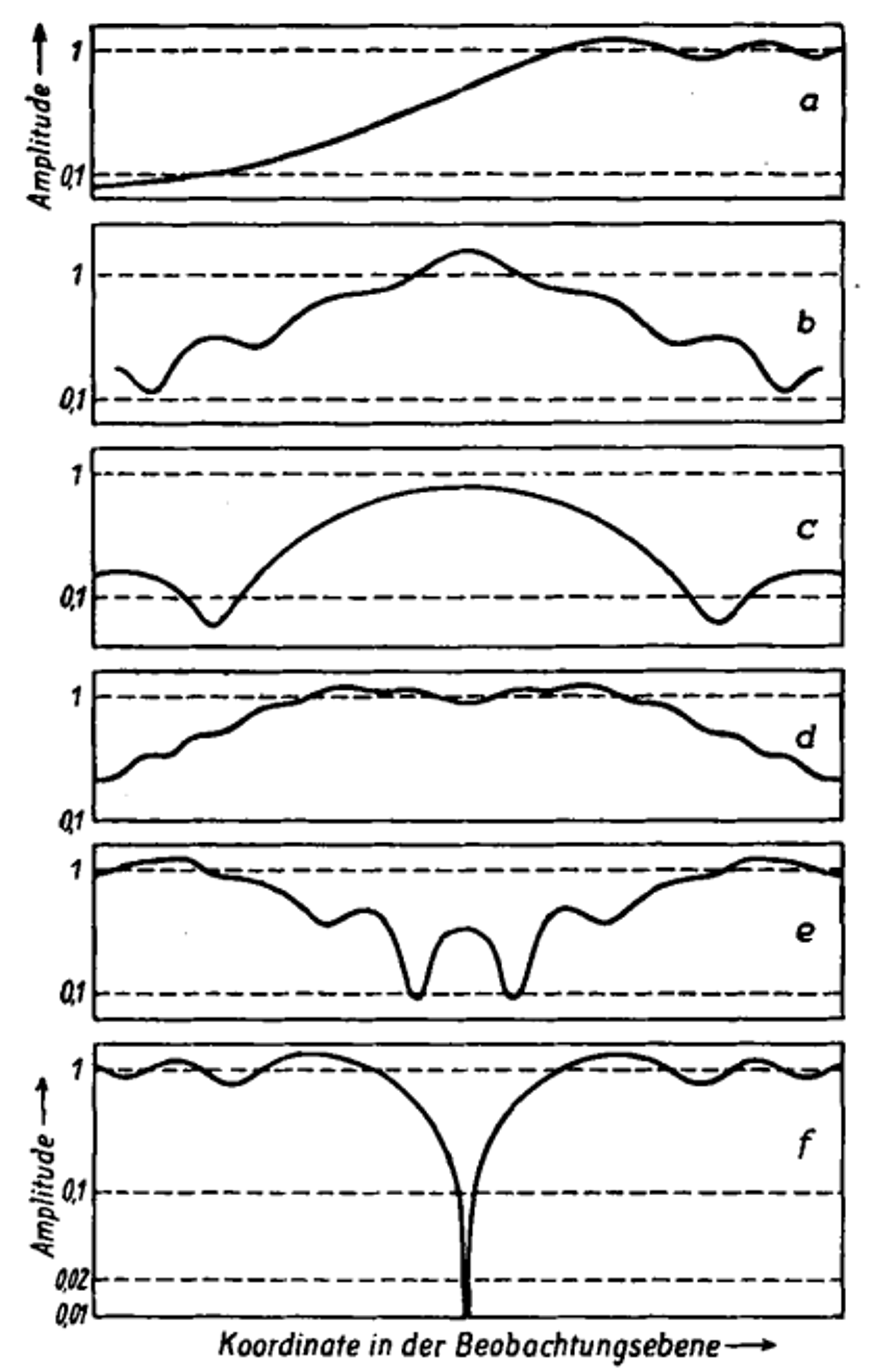}
		\includegraphics[width=0.49\linewidth]{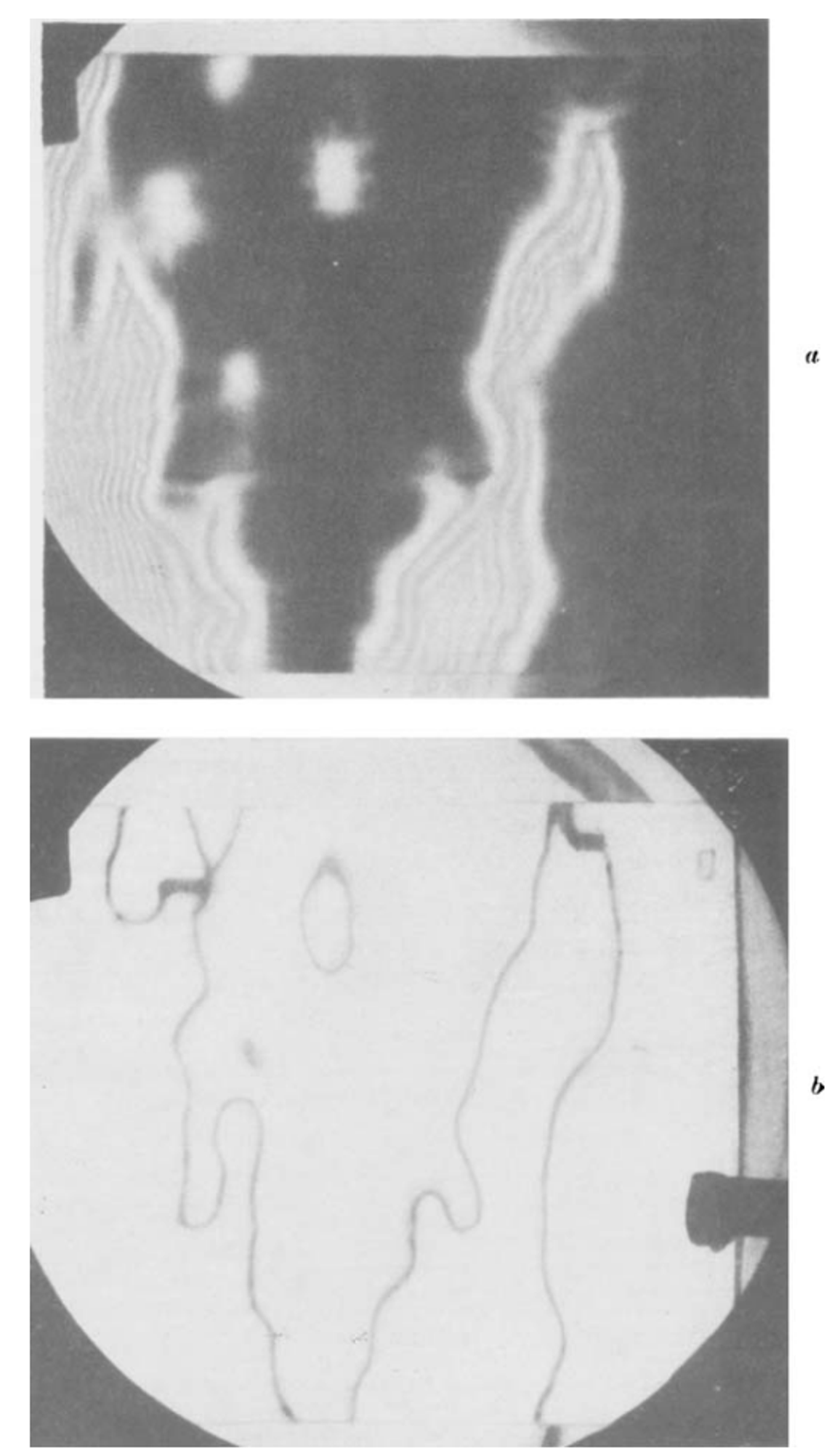}
		\caption{ }
		\begin{minipage}{11.4cm}
			\centering
			Left: Increase in resolution by minimum beam marking: between light source and plane of observation: a) opaque half-plane, b) slit of optimum width, c) and d) slits of smaller and larger widths. e) a wire of optimum thickness and f) half-wave plate for minimum beam marking. Right: Comparison of Toepler's schlieren photography (a) and minimum beam marking (b) of a glass plate [\cite{Wolter1950b}].
			\\
			
			\
		\end{minipage}
	\end{center}
\end{figure}

Interestingly the relation of Wolter's idea with the super resolving STED microscope of Stefan W. Hell (Nobel Prize in Chemistry 2014) has not yet been recognized. By marking the sample with a STED pulse, Hell created a circular interference field with a minimum in shape of a zero \emph{beam} of theoretically infinitesimal resolution [\cite{Hell1994}]. This is the same effect that Wolter used, with the only difference that in Wolter's microscope a zero \emph{plane} was marked.

\subsection{Imaging X-ray optics}
\label{sec_x-ray}

Thinking about enhancement of the resolution of microscopes for studying biological samples it is obvious to consider the use of electromagnetic waves of much shorter wavelengths than in the visible range. Wolter suggested to use 2.4 nm waves, i.e.\ X--rays, which are not absorbed by water but by carbon and nitrogen containing structures.

Focusing of X--rays by grazing--incidence reflection had been shown by Kirkpatrick and Baez
[\cite{Kirkpatrick1948}], but they used two orthogonal mirrors with only one reflection in any spatial direction, resulting in a non-aplanatic optical system of bad focusing properties. Wolter realized that a system obeying the Abbe criterion, i.d. free of coma and optical aberration, must be composed of two mirrors of different curvature. He suggested three different designs. Type I and type II make use of a paraboloid and a hyperboloid, type III combines a paraboloidal and an ellipsoidal mirror. In each case the two mirrors involved are arranged in a coaxial and confocal manner  [\cite{Wolter1952a}]. The main difference between the three types is the ratio of focal length to total system length, i.e.\ the minimum physical length of the optics.

\begin{figure}[h]
\begin{center}
	\includegraphics[width=10.5cm]{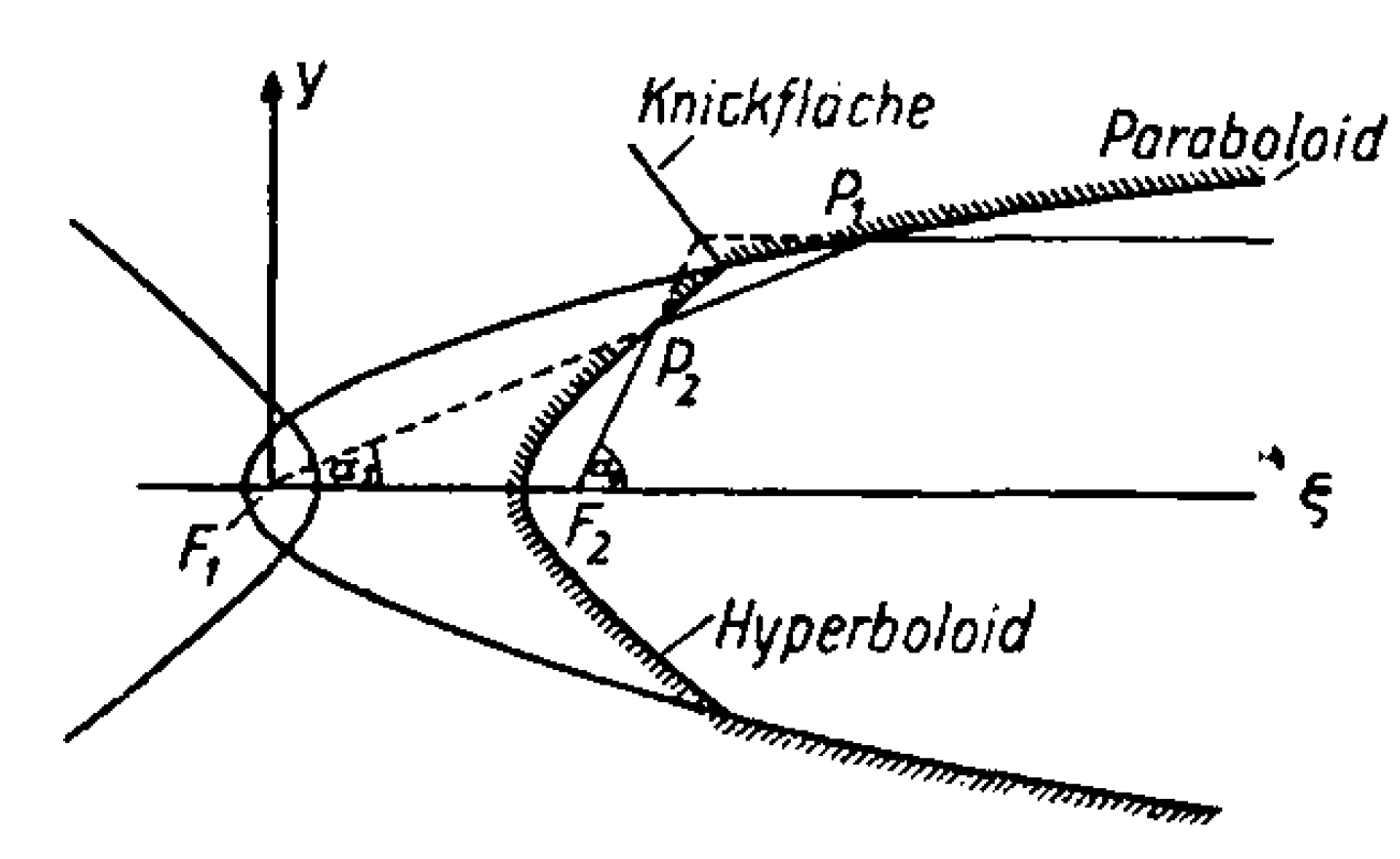}
	\caption{ }
	\begin{minipage}{11.5cm}
		\centering
	X--ray optics of Wolter's type I\\
	
	 [\cite{Wolter1952a}]. 
	\end{minipage}
\end{center}
\end{figure}

\begin{figure}[H]
\begin{center}
	\includegraphics[width=11.5cm]{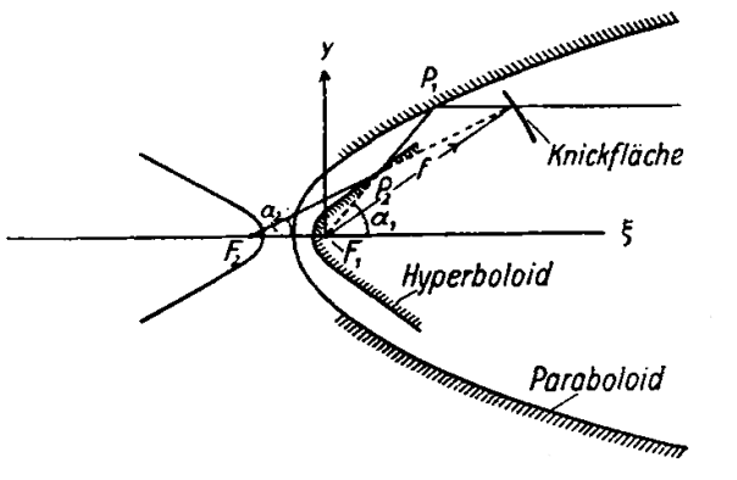}
	\includegraphics[width=11.5cm]{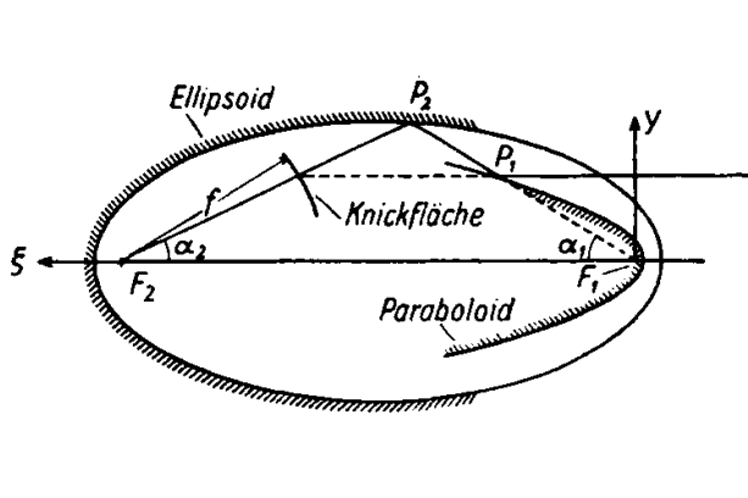}	
	\caption{ }
		\begin{minipage}{11.5cm}
		\centering
		X--ray optics of Wolter's types II (left) and III (right)\\
		
		 [\cite{Wolter1952a}].
	\end{minipage}
\end{center}
\end{figure}

Wolter-type systems are free of spherical aberration, but still suffer from
coma aberration, astigmatism and field curvature for larger field of view. In a second paper Wolter [\cite{Wolter1952b}] presented the equations for grazing--incidence optics which exactly obey the Abbe criterion. This is achieved by very small corrections of the axial mirror profile. The exact surface shape has been derived by Wolter by extending the solutions to grazing incidence which Karl Schwarzschild had already obtained for normal incidence in 1905 [\cite{Schwarzschild1905}]. Therefore these systems are named Wolter--Schwarzschild optics.

It turned out, that Wolter--optics are not suitable for X--ray microscopes, though in his book dated 1976 [\cite{Wolter1976}] Wolter still expressed this hope. However, these grazing--incidence reflection optics have shown to be superior over any other design for X--ray telescopes so far. To overcome the small field of view of a grazing--incidence reflection system, several interleaved concentric mirrors are used. This can only be realized by Wolter's type I and type II design, the mostly used design is type I.
The first imaging X--ray telescope, the Einstein Observatory (NASA), could take images with an angular resolution of 5 arcsec, the ROSAT (DLR/NASA) was designed for a resolution of 3 arcsec and the Chandra X--ray Observatory (NASA) is reaching 0.5 arcsec [\cite{Gorenstein2013}].

The Einstein Observatory was launched 3 months after Wolter's death. However,
Wolter must have realized the use of optics for telescopes. In 1971, almost 20 years after his first two papers about imaging X--ray optics he published a third paper concerning X--ray systems, where he estimated the imaging errors for his design to be used in telescopes [\cite{Wolter1971}].

\section{Conclusion}

Hans Wolter surely can be recognized as a pioneer of applied optics. 
Nowadays his ideas, his scientific results --- about 100 papers including two review articles in books --- are widely used in different areas, definitely with a huge impact in astronomy, too. And, some of his ideas have shown visionary sights, e.g.\ the relation of minimum beam marking to the STED microscope.

There is one good reason for his success: Wolter was combining theory and experiment on a very high level in one person!


\subsection{Acknowledgements}
	
My special thanks are addressed to Prof.\ Dr.\ Ulrich Gradmann, one of the former coworkers of Hans Wolter, for his insights in the time of Wolter in Marburg and especially for pointing out the connection between the minimum beam marking and the Nobel Prize in Chemistry to Stefan W. Hell in 2014. I like to thank Prof.\ Dr.\ Peter Thomas and Dr.\ Jürgen--Peter Schmidt, both former students of Hans Wolter, for many information about studying under Wolter. I appreciate the cooperation with the Archive of the Philipps-Universität Marburg. Especially Dr.\ Carsten Lind was very helpful in finding the documents concerning Hans Wolter and the Institute of Applied Physics in Marburg.



\end{document}